\documentclass[final]{aipproc}

\layoutstyle{8x11single}

\usepackage[utf8x]{inputenc}
\usepackage[english]{babel}
\usepackage{bm}
\usepackage{mathrsfs}
\usepackage{color}
\usepackage{xcolor}
\definecolor{magenta}{rgb}{1.2,0.15,0.6}
\definecolor{gold}{rgb}{1,0.8,0.3}
\definecolor{lightblue}{rgb}{0.5,0.8,1.6}
\definecolor{lightblue2}{rgb}{0.5,0.5,1.0}
\definecolor{dark}{rgb}{0.10,0.2,0.3}
\definecolor{light}{rgb}{1.7,1.5,0.6}
\definecolor{light2}{rgb}{1,0.9,0.3}
\definecolor{purpure}{rgb}{0.5,0.15,0.3}
\definecolor{bluelight}{rgb}{0.95,1.4,1.4}
\definecolor{fucsia}{rgb}{0.8,0.1,0.5}
\definecolor{violet}{rgb}{0.5,0.1,0.95}
\definecolor{bluegreen}{rgb}{0,0.6,0.6}
\usepackage{amsmath}
\usepackage{amssymb}

\begin{document}

\title{NLO Vertex for a Forward Jet plus a Rapidity Gap at High Energies}


\author{Martin Hentschinski}{
  address={Department of Physics, Brookhaven National Laboratory, Upton, NY 11973, USA.}
}

\author{José Daniel Madrigal Martínez\footnote{Speaker}~}{
  address={Institut de Physique Théorique, CEA Saclay, F-91191 Gif-sur-Yvette, France.}
}

\author{Beatrice Murdaca}{
  address={INFN, Grupo Collegiato di Cosenza, I-87036 Arcavacata di Rende,
Cosenza, Italy.}
}

\author{Agustín Sabio Vera}{
  address={Instituto de Física Teórica UAM/CSIC, Nicolás Cabrera 15 \\\& Facultad de Ciencias,
Universidad Autónoma de Madrid, C.U. Cantoblanco, E-28049 Madrid, Spain.}
  ,altaddress={CERN, Geneva, Switzerland.} 
}

\begin{abstract}
We present the calculation of the forward jet vertex associated to a rapidity gap (coupling of a hard pomeron to the jet) in the BFKL formalism at next-to-leading order (NLO). Real emission contributions are computed via Lipatov's effective action. The NLO jet vertex turns out to be finite within collinear factorization and allows, together with the NLO non-forward gluon Green's function, to perform NLO studies of jet production in diffractive events (e.g. Mueller-Tang dijets).
\end{abstract}

\maketitle

\section{Dijets with Rapidity Separation as a Probe of BFKL Pomeron}

More than twenty years ago, Mueller and Tang \cite{MuellerTang} proposed the study of the cross-section for dijet production with an associated large rapidity gap as an ideal environment to test the resummation effects associated to the BFKL pomeron \cite{BFKL}. In comparison to other observables that test the multi-Regge limit of QCD like angular decorrelation of Mueller-Navelet jets \cite{MuellerNavelet}, the growth of the proton structure functions with decreasing Bjorken-$x$ \cite{F2} or the total inclusive $\gamma^*\gamma^*$ cross-section \cite{photon}, hard exclusive diffraction processes with large momentum transfer are sensitive not only to the pomeron intercept but also to its slope, probing the BFKL Green's function at $t\ne 0$, currently known at NLO \cite{nonf}.\\

Next-to-leading logarithmic corrections are known to be sizeable in BFKL physics. This is expected to be the case at the level of the universal Green's function, i.e. the amplitude for pomeron exchange, since NLO corrections incorporate the dependence on the renormalization, factorization and reggeization scales which formally correspond to subleading terms in the LO approximation. However, it turns out that NLO corrections coming from the process-dependent impact factors (the couplings of the pomeron to the projectile and target) are also quite important (see, e.g. \cite{photon}). Therefore, in order to provide a reliable description of rapidity gap phenomenology within the BFKL approach, it is mandatory to go beyond the approximations used in previous works \cite{previous}, where NLO corrections were only included (partially) at the level of the Green's function.\\

A quantitative understanding of the Mueller-Tang cross-section faces a number of challenges. On the one hand, the configurations of interest with color singlet exchange, which do not generate any emission into the gap, compete with Sudakov-suppressed color exchange contributions where emissions are allowed up to a scale set by the experimental resolution $E_{\rm gap}$ of the rapidity gap definition. Moreover jet-gap-jet events are affected by soft rescatterings of the proton remnants which destroy the gap. We will focus on the singlet exchange configuration while assuming that the latter contribution can be taken into account through a rapidity gap survival factor. On the other hand, even at the level of the calculation of the singlet exchange contribution, we are faced with interesting and challenging questions, for instance: Does the exclusive nature of our observable preclude the applicability of collinear factorization? Moreover, does it make sense (at NLO) the Mueller-Tang prescription \cite{MuellerTang, Bartels} to couple the pomeron to a single quark, thus enabling the computation within collinear factorization? At the impact factor level, we find rather non-trivially that all collinear singularities in the jet are cancelled by DGLAP renormalization of the PDFs. In any case, a more conclusive answer to these questions can only be given after considering the convolution with the Green's function. A promising way to do this is the Monte-Carlo implementation of the NLO BFKL equation directly in ${\bm k}_T$ space already used in \cite{montecarlo}.

\section{High-Energy Factorization, Lipatov's Action \& Quasielastic Corrections}

\begin{figure}[ht]
\includegraphics[width=0.8\textwidth]{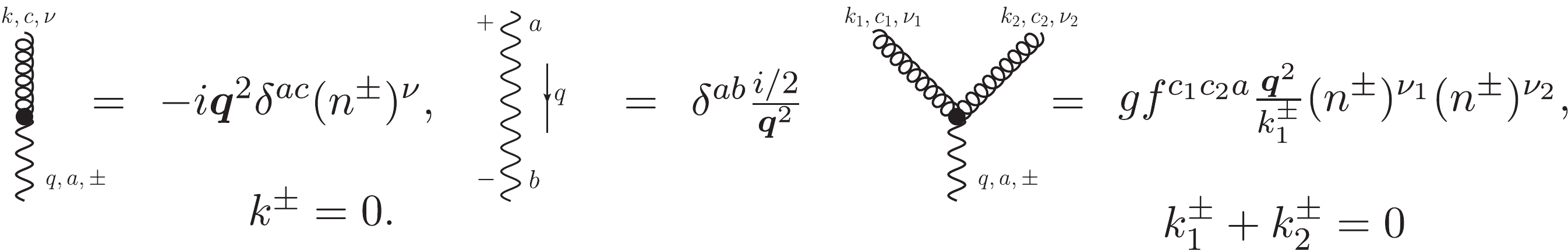}
\caption{Feynman rules for the lowest-order induced vertices of Lipatov's action \eqref{action}. Wavy lines denote reggeons, and curly lines, gluons.}
\label{induced}
\end{figure}

In multi-Regge kinematics, multi-particle production amplitudes are naturally factorized into a simple expression given by the product of \emph{reggeon} propagators and Lipatov effective vertices \cite{BFKL}. In a seminal work \cite{Levaction} (see also \cite{Levaction2}), Lipatov showed that this high-energy factorization can be preserved in a gauge-invariant fashion also when considering logarithmically- (non-power-) suppressed contributions in the high-energy limit (quasi-multi-Regge kinematics). To do this, it is necessary to introduce an infinite number of induced vertices that follow from the action
\begin{equation}\label{action}
 S_{\rm eff}=S_{\rm QCD}+S_{\rm ind}=\int{\rm d}^4x\,{\rm Tr}[(W_-[v(x)]-A_-(x))\partial_\perp^2 A_+(x)+\{+\leftrightarrow -\}],\quad \partial_\pm A_\mp(x)=0.
\end{equation}
Here $v_\mu=-{\rm i}T^a v^a_\mu(x)$ is the gluon field, $A_\pm(x)=-{\rm i}T^a A^a_\pm (x)$  the \emph{reggeon} field, and $W_\pm[v(x)]=-\frac{1}{g}\partial_\pm{\cal P}\exp\left\{\-\frac{g}{2}\int_{-\infty}^{x^\mp}{\rm d}z^\pm v_\pm (z)\right\}$, the Wilson line coupling whose perturbative expansion generates all induced vertices (Fig. \ref{induced}).\\

In the original formulation it is tacitly implied that the usual QCD interactions are restricted to the interactions of particles with similar rapidities, while non-local in rapidity interactions are mediated through reggeon exchange, so that no overcounting of degrees of freedom occurs. In recent years, we devised a manifestly gauge-invariant procedure to avoid this double-counting and make sense of the new appearing rapidity divergences when one considers amplitudes beyond tree level \cite{procedure1}, and used it to provide an independent computation of the 1-loop corrections to the forward jet vertex and the 2-loop gluon Regge trajectory \cite{procedure1,procedure2}.\\

\begin{figure}[th]
  \centering
\includegraphics[height = 2.2cm]{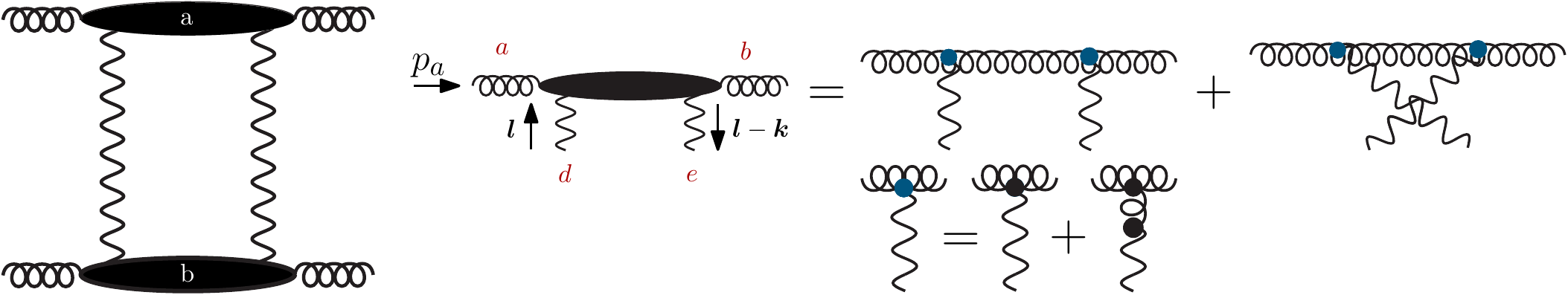}  
  \caption{{\em Left:} The leading log amplitude for gluon induced jets in the high-energy approximation. The state of two reggeized gluons in the $t$-channel is projected onto the color singlet; {\em Right:} Leading order diagrams which describe within the effective action the coupling of the gluon to the two reggeons.}
  \label{fig:schematic}
\end{figure}

Lipatov's action thus provides a highly efficient tool to compute radiative corrections for high-energy processes. In this framework, the (bare) pomeron appears as a state of two reggeons projected into a color singlet state (Fig.~\ref{fig:schematic}). High-energy factorization suggests absorbing in the impact factor definition the integral over longitudinal components of the reggeon loop momenta; therefore, after BFKL resummation of $\Delta y_{\rm gap}\sim\ln (\hat{s}/s_0)$ enhanced terms, we have
\begin{equation}\label{full}
\frac{{\rm d}\hat{\sigma}^{\rm res}_{ij}}{{\rm d}^2{\bm k}}=\int\frac{{\rm d}^2{\bm l}_1}{\pi}\int{\rm d}^2{\bm l}_1'G({\bm l}_1,{\bm l}_1',{\bm k},\hat{s}/s_0)\int\frac{{\rm d}^2{\bm l}_2}{\pi}\int{\rm d}^2{\bm l}_2'G({\bm l}_2,{\bm l}_2',{\bm k},\hat{s}/s_0)h_{i,{\rm a}}h_{j,{\rm b}},\quad h_i=h_i^{(0)}+h_i^{(1)}.
\end{equation}
Here $\hat{s}$ is the partonic c.o.m. energy squared, ${\bm k}^2$ the momentum transfer and $G$ the Green's function. We assume that, like in the  LO case \cite{motyka}, the expression \eqref{full} after full resummation encounters no divergences in the integrations over the transverse loop momenta in the direct (${\bm l}_1,{\bm l}_1'$) and complex conjugate (${\bm l}_2,{\bm l}_2'$) amplitudes. At LO, the computation of the impact factor (in $d=2+2\epsilon$) is very simple (Fig.~\ref{fig:schematic}), and in the gluon-initiated case we have
\begin{equation}
 {\rm i}\phi_{gg}=\int\frac{{\rm d}l^-}{8\pi}{\rm i}{\cal M}^{abde}_{gr^*r^*\to g}P^{de},~\quad P^{de}=\frac{\delta^{de}}{\sqrt{N_c^2-1}}~~~~\Longrightarrow~~~~h_g^{(0)}=\frac{\overline{|\phi_{gg}|^2}}{2(8\pi^2)^{1+\epsilon}(p_a^+)^2}=h^{(0)}(1+\epsilon)C_a^2,
\end{equation}
with $h^{(0)}=\frac{\alpha_{s,\epsilon}^2 2^\epsilon}{\mu^{4\epsilon}\Gamma^2(1-\epsilon)(N_c^2-1)}$ and $\alpha_{s,\epsilon}=\frac{g^2\mu^{2\epsilon}\Gamma(1-\epsilon)}{(4\pi)^{1+\epsilon}}$. In a similar way, in the quark-initiated case we have $h_q^{(0)}=C_f^2h^{(0)}$.\\
\begin{figure}[th]
  \centering
\includegraphics[scale=0.6]{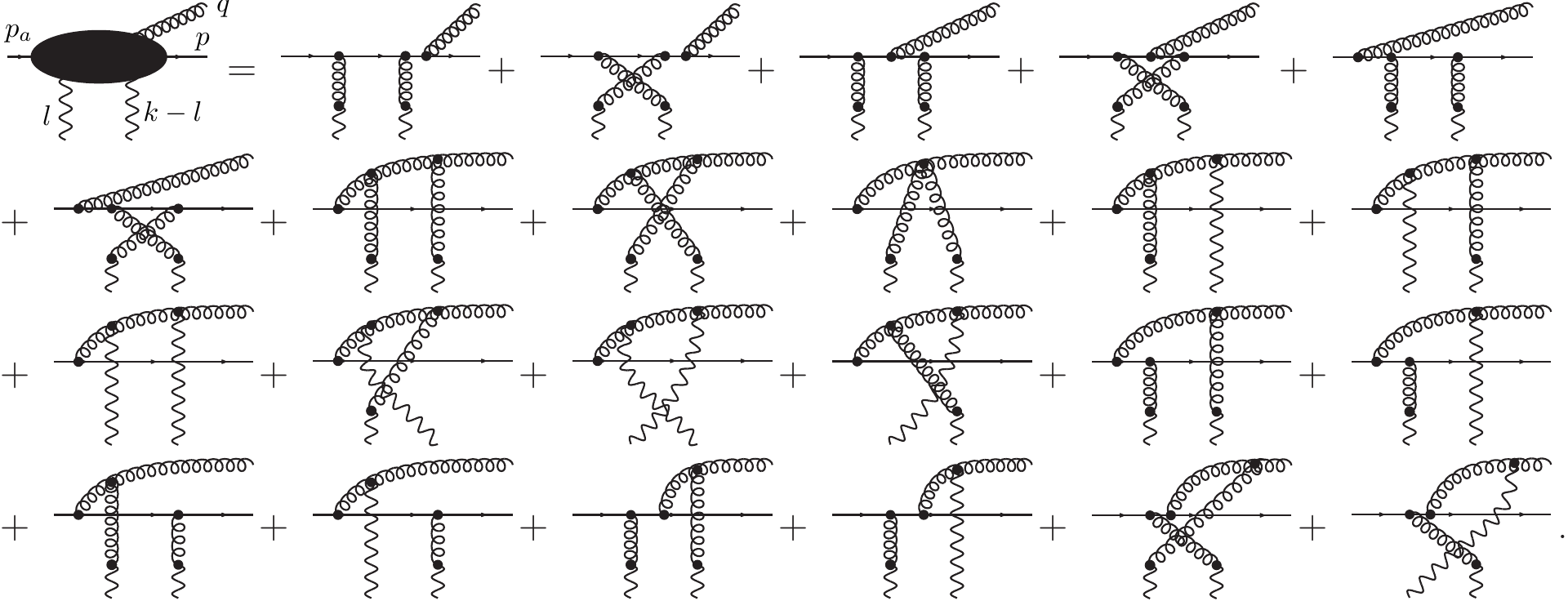}  
  \caption{Effective action diagrams for the quasielastic emission corrections in the  $q(\bar{q})\to q(\bar{q})g$ channel.}
  \label{quasi}
\end{figure}

At NLO, the computation is much more involved. Virtual corrections were already computed in \cite{virtual}. The missing ingredient is the evaluation of the quasielastic emission corrections, like those appearing in Fig. \ref{quasi} for the $q(\bar{q})\to q(\bar{q})g$ channel. Similarly, we must also compute the diagrams corresponding to the $g\to gg$ and $g\to q\bar{q}$ channels. At the level of the differential partonic impact factor, very compact expressions are found \cite{tang}:
\begin{equation}\label{diffac}
\parbox{18cm}{ \includegraphics[scale=0.52]{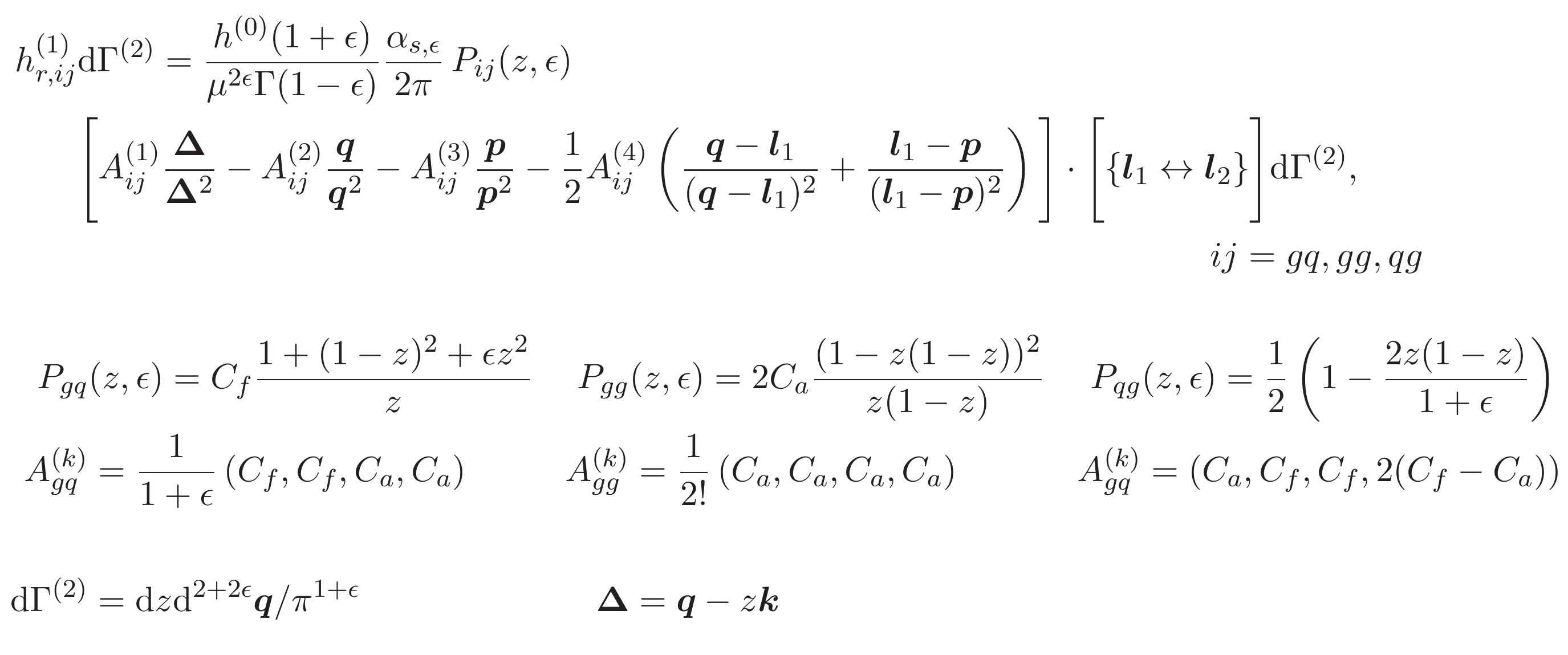}}
\end{equation}
where $z$ (resp. $1-z$) is the longitudinal momentum fraction carried by the outgoing parton with momentum $q$ ($p$).
\section{The NLO Mueller-Tang Jet Vertex}
The phase space integration in \eqref{diffac} reveals singularities that are manifested as poles in the dimensional regularization parameter $\epsilon$. As usual, in order to define an IR and collinear safe cross-section, we need to convolute the partonic cross-section with a jet function distribution $S_J$ \cite{catani}: $\frac{{\rm d}\hat{\sigma}_J}{{\rm d}J_1{\rm d} J_2{\rm d}^{2+2\epsilon}{\bm k}}={\rm d}\hat{\sigma}\otimes S_{J_1}S_{J_2}, {\rm d}J_i={\rm d}^{2+2\epsilon}{\bm k}_{J_i}{\rm d}y_{J_i},\,i=1,2.$\\

For a general jet algorithm, one can then prove that, by using an explicit phase slicing parameter $\lambda$ which isolates the singular regions in the integrations, one obtains a finite expression for the NLO corrections to the effective jet vertex $\frac{{\rm d}\hat{V}^{(1)}}{{\rm d}J}$ in $d=4$ after including the virtual corrections and taking into account the renormalization of the coupling and the DGLAP renormalization of the parton densities \cite{tang}. Then, it is possible to write within collinear factorization
\begin{equation}
\begin{aligned}
\frac{{\rm d}\sigma_{J,H_1H_2}}{{\rm d}J_1{\rm d}J_2{\rm d}^2{\bm k}}&=\frac{1}{\pi^2}\int{\rm d}{\bm l}_1{\rm d}{\bm l}_1'{\rm d}{\bm l}_2{\rm d}{\bm l}_2'\frac{{\rm d}V({\bm l}_1,{\bm l}_2,{\bm k},{\bm p}_{J,1}, y_1,s_0)}{{\rm d} J_1}G({\bm l}_1,{\bm l}_1', {\bm k},\hat{s}/s_0)G({\bm l}_2,{\bm l}_2', {\bm k},\hat{s}/s_0)\frac{{\rm d}V({\bm l}_1',{\bm l}_2',{\bm k},{\bm p}_{J,2}, y_2,s_0)}{{\rm d} J_2},\\
\frac{{\rm d}V}{{\rm d}J}&=\sum_{j=\{q,\bar{q},g\}}\int_{x_0}^1{\rm d}x\,f_{j/H}(x,\mu_F^2)\left(\frac{{\rm d}\hat{V}_j^{(0)}}{{\rm d}J}+\frac{{\rm d}\hat{V}_j^{(1)}}{{\rm d}J}\right),\quad x_0=\frac{{\bm k}^2}{M_{X,{\rm max}}^2+{\bm k}^2}.
\end{aligned}
\end{equation}
We refer the reader to \cite{tang} for the final expressions for $\frac{{\rm d}\hat{V}^{(1)}}{{\rm d}J}$. We plan to carry out in the near future the numerical implementation of this vertex which, as already remarked, is a necessary step towards reliable phenomenological studies of diffractive jet production at high energies. 
\vspace{-0.1cm}
\begin{theacknowledgments}
 We would like to thank the organizers of DIFFRACTION 2014 for the oportunity to present our research in such a pleasant environment. M.H. acknowledges support from U.S. Department of Energy (DE-AC02-98CH10886) and BNL grant LDRD 12-034. J.D.M. is supported by Advanced Investigator Grant ERC-AD-267258. A.S.V. acknowledges support from European Commission under contract LHCPhenoNet (PITN-GA-2010-264564), Madrid Regional Government (HEPHACOSESP-1473), Spanish Government (MICINN (FPA2010-17747)) and Spanish MINECO Centro de Excelencia Severo Ochoa Programme (SEV-2012-0249).
\end{theacknowledgments}

\end{document}